\documentclass[a4paper,twocolumn]{revtex4}
\usepackage{graphicx,dcolumn,bm,amsmath}
\usepackage[normalem]{ulem}

\usepackage[dvipsnames,table]{xcolor}
\usepackage{color}

\newcommand{\beq}{\begin{equation}}
\newcommand{\enq}{\end{equation}}

\newcommand\I{\mbox{i}}
\newcommand{\e}{\eqref}

\newcommand \bfr{{\bf r}}
\newcommand \bfk{{\bf k}}

\newcommand{\p}{\partial}

\begin{document}

\title{Explosive Development of the Quantum Kelvin-Helmholtz Instability on the He-II Free Surface}


\author{N. M. Zubarev$^{1,2}$ and P. M. Lushnikov$^{3,4}$}

\address{
$^1$Institute of Electrophysics, Ural Branch, Russian Academy of Sciences, Yekaterinburg, 620016 Russia \\
$^2$Lebedev Physical Institute, Russian Academy of Sciences, Moscow, 119991 Russia \\
$^3$Landau Institute for Theoretical Physics, Russian Academy of Sciences, Chernogolovka, Moscow oblast, 142432 Russia  \\
$^4$Department of Mathematics and Statistics, University of New Mexico, Albuquerque, 87131-0001 NM, USA  \\
}

\begin{abstract}
We analyze nonlinear dynamics of the Kelvin-Helmholtz quantum instability
of the He-II free surface, which evolves during counterpropagation of the
normal and superfluid components of liquid helium. It is shown that in the
vicinity of the linear stability threshold, the evolution of the boundary
is described by the $|\phi|^4$ Klein-Gordon equation for the complex
amplitude of the excited wave with cubic nonlinearity. It is important
that for any ratio of the densities of the helium component, the
nonlinearity plays a destabilizing role, accelerating the linear
instability evolution of the boundary. The conditions for explosive growth
of perturbations of the free surface are formulated using the integral
inequality approach. Analogy between the Kelvin-Helmholtz quantum
instability and electrohydrodynamic instability of the free surface of
liquid helium charged by electrons is considered.

\end{abstract}

\maketitle

\section{ INTRODUCTION }

\begin{figure}[h]
\center{\includegraphics[width=0.8\linewidth]{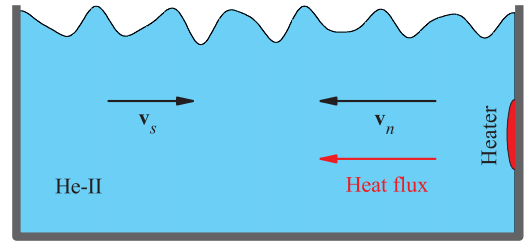}}
\caption{
Counterpropagation (with velocities  ${\bf v}_{s,n}$) of the superfluid and normal components of superfluid $^4$He induced by a heat flow from a heater, which is carried by the normal component. Both components are located in the same volume of  $^4$He and flow along the tangent to the common free surface.} \label{ref:KHIschematic}
\end{figure}

The tangential discontinuity of velocities in a liquid as well as at the
interface between two liquids leads to the emergence of the classical
Kelvin-Helmholtz instability (KHI) \cite{LandauLifshitzHydrodynamics1989}.
In recent years, the KHI in superfluid liquids has been actively studied,
including the instability of the interface between different superfluid
phases of  $^3$He
\cite{VolovikJETPLett2002,BlaauwgeersEltsovVolovikEtAlPRL2002,VolovikBook2003,
FinneEltsovKopninVolovikEtAlRepProgrPhys2006, VolovikUspekhiFizNauk2015}
as well as the instability of free  $^4$He surface
\cite{HanninenBaggaleyPNAS2014,RemizovLevchenkoMezhovDeglinJLowTempPhys2016,BabuinLvovEtAlPRB2016}
 in the superfluid state (the so-called He-II phase appearing at a temperature below  2.17~K  \cite{KhalatnikovBook1971}). The former case realized for
$^3$He, is the closest to the classical KHI (the phases are on different
sides of the boundary), while the latter case is principally different
since instability appears due to counterpropagation of the normal and
superfluid  $^4$He components under the free surface. In this study, we
consider the second case that can naturally be referred to as quantum KHI
since both components are on the same side of the free surface, and their
coexistence is a quantum effect having no classical analog. A typical
experimental situation, in which such a relative motion of components is
observed, is illustrated in  Fig.~\ref{ref:KHIschematic}. Experimental
investigations of the emergence of instability on the free flat surface of
superfluid He-II with the heat flow in the bulk of the liquid have been
initiated by I.M. Khalatnikov and are actively performed at the Laboratory
of Quantum Crystals, Institute of Solid State Physics, Russian Academy of
Sciences (see, for example,
\cite{AbdurahimovLevchenkoMezhovDeglinKhalatnikovLowTempPhys2012,RemizovLevchenkoMezhovDeglinJLowTempPhys2016,LevchenkoMezhovDeglinPelmenevJETPLett2017,LevchenkoMezhovDeglinPelmenevFNT2018,LevchenkoMezhovDeglinPelmenevMaterialsLetters2019}.

We will use the two-liquid approximation for describing the dynamics of
$^4$He  \cite{KhalatnikovBook1971} with densities $\rho_s$ and $\rho_n$ of
the superfluid and normal components, respectively (total density of the
liquid is  $\rho =\rho _n + \rho _s $).   Both components are treated as
incompressible liquids  ($\rho _n = \mbox{const}$ and
$\rho_s=\mbox{const})$.

Quantum KHI instability increment for linear perturbations proportional to  $\propto e^{\I \bfk \cdot \bfr_{\!\bot}-\I\omega t}$ of a flat horizontal surface in the presence of gravity and capillarity is
given by the following dispersion relation obtained in
\cite{KorshunovEuriphysLett1991,KorshunovJETPLett2002}:
\begin{align} \label{KHIdispersionKorshunov}
\omega_k^2=\rho_s(\omega-{\bf V}_s{ \cdot \bfk})^2+\rho_n\left (\omega-{\bf V}_n\cdot\bfk+\I2\nu_nk ^2\right )^2\nonumber \\+4\rho_n\nu_n^2k^3m_n,
\end{align}
where ${\bf V}_s$ and ${\bf V}_n$ are the mean velocities of the superfluid and normal components,  $\bfr_ {\!\bot}$  is the horizontal coordinate;  $t$ is the time, $\nu_n$ is the kinematic viscosity of the normal component, which is defined as the dynamic viscosity normalized to $\rho_n$; $\bfk$ and $\omega$ are the wavevector and frequency of perturbations,
$k=|\bfk|$ and $m_n=\left[ k^2-{\I(\omega- {\bf V}_n{ \cdot \bfk})/\nu_n}{}\right]^{1/2}$.  Also
\begin{align} \label{omegakhelium}
\omega_k^2\equiv gk+\alpha  k^3/\rho
\end{align}
is the dispersion relation for gravity-capillary waves in the absence of
average motion of the liquid components, where $g$ is the acceleration due
to gravity and  $\alpha$ is the surface tension coefficient.

In the framework of the simplest nondissipative two-liquid description
\cite{KorshunovEuriphysLett1991,KorshunovJETPLett2002}  the flow of both
phases is treated as a potential flow. In this case, the velocities of the
phases can be written as ${\bf v}_{n,s} = \nabla \Phi_{n,s}$, where
$\Phi_{n,s}$ are the velocity potentials satisfying (since $\rho _n =
\mbox{const}$ and $\rho _s=\mbox{const}$) the Laplace equations
\begin{equation}
\label{eq1}
\nabla ^2\Phi _n = 0,\quad \quad \nabla ^2\Phi _s = 0,
\end{equation}
and dispersion relation \e{KHIdispersionKorshunov} can be reduced to
\begin{align} \label{KHIdispersionKorshunovpotential}
(\omega-{\bf V}_{m}{ \cdot \bfk})^2=\omega_k^2-\frac{\rho_s\rho_n}{\rho}({\bf V}{ \cdot \bfk})^2,
\end{align}
where ${\bf V}_m\equiv( \rho _n V_n + \rho _s V_s )/\rho$ is the mean velocity of the center of mass of the liquid and ${\bf V}\equiv{\bf V}_s-{\bf V}_n$ is the average relative velocity for the liquid components. Without loss of generality, we will henceforth assume that ${\bf V}_m=0$, which implies a transition to the corresponding moving frame of reference.

Dispersion relation \e{KHIdispersionKorshunovpotential} makes it possible
to find \cite{KorshunovEuriphysLett1991,KorshunovJETPLett2002} the
threshold value of relative velocity
\begin{align} \label{Vcdef}
V_c= \left( {\frac{4\rho^3g\alpha }{\rho _n^2 \rho _s^2 }} \right)^{1 / 4},
\end{align}
which corresponds to wavenumber  $k=k_0\equiv\sqrt{\rho g/\alpha}$.
A linear KHI appears at relative velocity  $V\equiv |{\bf V}|>V_c$.

It should be noted that the dispersion relation for the quantum KHI
assumes the same form as the conventional dispersion relation of the KHI
for the interface between two ideal immiscible liquids (see, for example,
\cite{LandauLifshitzHydrodynamics1989}), if we use in Eq.
\e{KHIdispersionKorshunovpotential} classical dispersion relation
$$\omega_{k,classical}^2=(\rho_s-\rho_n)gk/\rho+\alpha  k^3/\rho$$ for
gravity-capillary waves instead of relation \e{omegakhelium} for
$\omega_k$ in the absence of average motion of the liquid components (in
this case, we presume that the liquid of density $\rho_n$ is above the
interface, while the liquid with density  $\rho_s$ is below it). To return
from $\omega_{k,classical}$ to relation \e{omegakhelium},  it is
sufficient to perform substitution $\rho_n\to-\rho_n$  (without changing
total density $\rho$) in  $\omega_{k,classical}$. The remaining terms in
relation  \e{KHIdispersionKorshunovpotential}  are independent of $g$ and,
hence, do not change depending on the position of the second liquid with
density $\rho_n$ under the free surface (quantum case) or above the
interface (classical case). The described difference between the linear
dispersion relations for quantum and classical KHIs are only quantitative
by nature. A qualitative difference between the classical and quantum KHIs
appears at nonlinear stages of instability development. For example, in
the limit  $V\gg V_c$ for classical KHI, a tendency to the formation of
weak root singularities appears at the interface between the liquids, for
which the surface remains smooth, but its curvature becomes infinitely
large over a finite time interval
\cite{MooreProcRSocLond1979,Zubarev_Kuznetsov_JETP_2014}. Under analogous
conditions for the quantum KHI, a tendency to the formation of strong
singularities (cusp points) appears \cite{LushnikovZubarevPRL2018}.

In this study, we consider nonlinear stages of development of the quantum KHI in the vicinity of the stability threshold (i.e., for $|V-V_c|/V_c\ll 1$). In this situation, a narrow packet of surface waves in the Fourier space is excited, which makes it possible to formulate the equation of the envelope of this wave packet. It will be shown that the nonlinearity for any relation between the densities of helium components produces a destabilizing effect, i.e., accelerates the development of the linear instability of the interface and leads to explosive instability with a singularity in the equation of the envelope appearing over a finite time interval. In the context of complete hydrodynamic equations, this means that the solution becomes strongly nonlinear (values of characteristics slopes become of the order of unity) over a finite time.

The article is organized as follows. In Section \ref{sec:basicequations}, we
consider basic equations in two-liquid hydrodynamics with kinematic and
dynamic boundary conditions on the free surface. In Section
\ref{sec:singlfluid}, a transformation to the effective one-liquid
description is made for 2D flows using harmonically conjugate potentials
(stream functions). In Section \ref{sec:Helimumlectric},  we demonstrate
that as a result of our transformations, the initial problem of
description of nonlinear development of the quantum KHI, which appears due
to relative motion of the normal and superfluid phases, becomes
equivalent (up to trivial removal of constants) to the problem of dynamics
of the electron-charged boundary of liquid helium in an electric field
(the limit when the charge is completely screens the fields over the
liquid). This analogy has allowed us to use a number of results obtained
earlier from analysis of the behavior of liquid helium in an electric
field for the problem considered here. In Section \ref{sec:amplitudeeq},
using the results obtained in
\cite{Gorkov_Chernikova_FNT_1976,Gorkov_Chernikova_DAN_1976}, we
demonstrate that the evolution of the interface in the vicinity of the
stability threshold can be described by the $|\phi|^4$ relativistically
invariant Klein-Gordon equation with nonlinear attraction for the complex
envelope of the wave being excited. It is important that for any relation
between the densities of the helium components, nonlinearity plays a
destabilizing role, accelerating the linear instability development. In
Section \ref{sec:sufficient}, using the integral inequality approach in
the Klein-Gordon equations and analogy with the motion of an effective
Newtonian particle in a certain potential, we formulate sufficient
conditions for explosive buildup of perturbations of the free surface. In
concluding Section \ref{sec:conclusion}, we consider the hard excitation
of strongly nonlinear solutions and their applicability in full two-liquid
hydrodynamics.

\section{BASIC EQUATIONS}
\label{sec:basicequations}

We limit our analysis to 2D flows for which all quantities depend on pair of variables  ${\bf r} = (x,y)$, where $x$ and $y$ are the horizontal and vertical coordinates, respectively. In this case, we have
$\nabla = \left(\partial/\partial x ,\partial/\partial y \right)$.

The helium surface in the unperturbed state is plane  $y = 0$, and the motion of helium components along the $x$ axis is uniform (i.e., equality $\Phi_{n,s}=V_{n,s} x$, holds for the velocity potentials, where $V_{n,s}$ are horizontal velocity components). As mentioned above, we can assume without loss of generality that $\rho _n V_n + \rho _s V_s = 0$, which corresponds to analysis of the problem in the center-of-mass system. Then the velocity components can be expressed in terms of average relative velocity $V=V_s-V_n>0$ of the components as  $V_{n,s} = \mp \rho _{s,n} V /\rho $.

We assume that the perturbed free surface of liquid helium is defined by equation $y = \eta (x,t)$; i.e., the liquid occupies domain
\[
-\infty < x < \infty , \qquad -\infty < y < \eta (x,t).
\]
Perturbations of velocity potentials, which appear because of deformation of the boundary, decay in the bulk:
\begin{equation}
\label{eq2}
\Phi_{n,s}\to V_{n,s} x, \qquad y \to - \infty.
\end{equation}
The motion of the boundary is determined by the dynamic and kinematic boundary conditions. The dynamic condition (time-dependent Bernoulli equation for a two-component liquid) has form
\begin{eqnarray}
\nonumber
\rho _n \left( {\frac{\partial \Phi _n }{\partial t} + \frac{(\nabla \Phi _n
)^2}{2}} \right) + \rho _s \left( {\frac{\partial \Phi _s }{\partial t} +
\frac{(\nabla \Phi _s )^2}{2}} \right)
\\
\label{eq3}
= - \rho g\eta + \frac{\alpha \eta
_{xx} }{(1 + \eta _x^2 )^{3 / 2}} + \Gamma ,\quad \quad y = \eta ,
\end{eqnarray}
where $\eta _x \equiv \partial \eta / \partial x$, $\eta _{xx} \equiv \partial
^2\eta / \partial x^2$; the first term on the right-hand side is responsible for the force of gravity, and the second term, for capillary forces. These forces tend to return the perturbed boundary of the liquid to the initial planar state. Quantity $\Gamma $ is the Bernoulli constant; its value that ensures the fulfillment of condition (\ref{eq3}) in the unperturbed state
$\Phi _{n,s} = V_{n,s} x$ and $\eta = 0$ is given by
\[
\Gamma = \frac{\rho _n V_n^2 + \rho _s V_s^2 }{2} = \frac{\rho _n \rho _s
V^2}{2\rho }.
\]
Finally, in accordance with the kinematic condition, normal velocity of the boundary must coincide with the normal component of the velocity in each phase,
\begin{equation}
\label{eq4}
\frac{\eta _t }{\sqrt {1 + \eta _x^2 } } = \partial _n \Phi _n = \partial _n
\Phi _s ,\quad \quad y = \eta (x,t),
\end{equation}
where $\eta _t \equiv \partial \eta / \partial t$,
and $\partial _n \equiv {\bf n} \cdot \nabla $
indicates the derivative with respect to the outward normal
 \[
{\bf n}= \left(-\eta _x ,1 \right)\frac{1}{\sqrt{1 + \eta_x^2}}
\]
to the boundary of the liquid.

\section{TRANSITION TO EFFECTIVE ONE-LIQUID DESCRIPTION}
\label{sec:singlfluid}

Let us introduce the average velocity of the medium as
 \begin{equation}
\label{eq5}
{\bf v} \equiv \frac{\rho _n {\bf v}_n + \rho _s {\bf v}_s }{\rho }.
\end{equation}
The equations of motion can be written in terms of a single effective liquid of density  $\rho $, which flows at velocity  ${\bf v}$. Velocity (\ref{eq5}) corresponds to potential
\begin{equation}
\label{eq6}
\Phi = \frac{\rho _n \Phi _n + \rho _s \Phi _s }{\rho },
\end{equation}
i.e.,  ${\bf v} = \nabla \Phi $. We also introduce auxiliary velocity potential
\begin{equation}
\label{eq7}
\phi = \sqrt {\rho _s \rho _n } (\Phi _n - \Phi _s ) / \rho.
\end{equation}
Potentials $\Phi $ and $\phi $ are linear combinations of harmonic potentials  $\Phi _{s,n} $; therefore, these potentials satisfy Laplace equations
\[
\nabla ^2\Phi = 0,\quad \quad \nabla ^2\phi = 0.
\]
Conditions (\ref{eq2}) deep inside fluid can be written as
\begin{equation}
\label{eq8}
\Phi \to 0,\quad \quad \phi \to - Vx\sqrt {\rho _s \rho _n } / \rho ,\quad
\quad y \to - \infty.
\end{equation}
Dynamic boundary condition  (\ref{eq3}) assumes the form
\begin{eqnarray}
\nonumber
\rho \left( {\frac{\partial \Phi }{\partial t} + \frac{(\nabla \Phi )^2}{2}}
\right) = - \rho g\eta + \frac{\alpha \eta _{xx} }{(1 + \eta _x^2 )^{3 / 2}}
\\
\label{eq9}
+ \Gamma - \frac{\rho (\nabla \phi )^2}{2},\quad \quad y = \eta .
\end{eqnarray}
It can easily be seen from expression (\ref{eq4}) that the kinematic
condition for potential  $\Phi$ becomes
\begin{equation}
\label{eq10}
\frac{\eta _t }{\sqrt {1+\eta _x^2}}=\partial _n \Phi ,\qquad y =\eta.
\end{equation}
Finally, the kinematic condition for potential $\phi$ is obviously trivial:
\begin{equation}
\label{eq11}
\partial _n \phi = 0,\quad \quad y = \eta .
\end{equation}
Thus, the initial equations of motion for the two components of liquid
helium in the nondissipative approximation can be reduced to classical
equations for a potential flow of a single incompressible liquid with a
free surface, which takes into account the capillary and gravity forces,
with additional term $\rho (\nabla \phi )^2/2$ on the right-hand side of
time-dependent Bernoulli equation  (\ref{eq9}). This term is responsible
for the effect of counterpropagation of the liquid helium componentsand,
hence, for the evolution of the Kelvin-Helmholtz instability. Let us
consider this term in greater detail.

We introduce auxiliary function $\psi $ which is the harmonic conjugate
with potential $\phi $, i.e., $\psi $ and $\phi $ are connected by the
Cauchy-Riemann relations
\[
\frac{\partial \phi }{\partial x} = \frac{\partial \psi }{\partial y},\quad
\quad \frac{\partial \phi }{\partial y} = - \frac{\partial \psi }{\partial
x}.
\]
Let us clarify the physical meaning of quantity $\psi $. Similarly to relation  (\ref{eq7}), it can be written as
\begin{equation}
\label{eq12}
\psi = \sqrt {\rho _s \rho _n } (\Psi _n - \Psi _s ) / \rho ,
\end{equation}
where $\Psi _{n,s}$ is the stream function for the normal and superfluid He-II components, which are connected with potentials $\Phi _{n,s}$ by relations
\[
\frac{\partial \Phi _{n,s} }{\partial x} = \frac{\partial \Psi _{n,s}
}{\partial y},\quad \quad \frac{\partial \Phi _{n,s} }{\partial y} = -
\frac{\partial \Psi _{n,s} }{\partial x}.
\]
Therefore, harmonic function $\psi $ is (to within a constant factor) the difference between the stream functions for different helium components.

As a consequence of the Cauchy-Riemann relations, we obtain $\left.
{\partial _n \phi } \right|_{y = \eta } = - \left. {\partial _\tau \psi }
\right|_{y = \eta } $, where
\[
\partial _\tau \equiv \frac{1}{\sqrt{1 + \eta_x^2}}
\left(1,\eta_x\right) \cdot \nabla
\]
is the tangential derivative. Then boundary condition (\ref{eq11}) can be written as
\[
\partial _\tau \psi = 0,\quad \quad y = \eta ,
\]
i.e., quantity $\psi $ does not change along the boundary. Without loss of generality, we can set  $\left. \psi \right|_{y = \eta } = 0$.

After the introduction of $\psi $, the equations of motion assume final form
\begin{equation}
\label{system1}
\nabla ^2\Phi = 0,\quad \quad \nabla ^2\psi = 0,
\end{equation}
\begin{equation}
\label{system2}
\Phi \to 0,\quad \quad \psi \to - Vy\sqrt {\rho_s \rho_n } / \rho ,\quad
\quad y \to - \infty ,
\end{equation}
\begin{eqnarray}
\nonumber
\rho \left( {\frac{\partial \Phi }{\partial t} + \frac{(\nabla \Phi )^2}{2}}
\right) = - \rho g\eta + \frac{\alpha \eta _{xx} }{(1 + \eta _x^2 )^{3 / 2}}
\\
\label{system3}
+ \Gamma - \frac{\rho (\nabla \psi )^2}{2},\quad \quad y = \eta ,
\end{eqnarray}
\begin{equation}
\label{system4}
\frac{\eta _t }{\sqrt {1 + \eta _x^2 } } = \partial _n \Phi ,\quad \quad
\psi = 0,\quad \quad y = \eta .
\end{equation}
It is important that the problem of determining function $\psi $ and, as a consequence, of key term $\rho (\nabla \psi )^2 / 2$ in the dynamic boundary condition decouples from the general problem of motion of the boundary. Indeed, it can be seen from these equations that quantity $\psi $ is completely determined by shape $\eta $ of the boundary and is independent of the velocity distribution (i.e., of potential $\Phi$). It is exactly this circumstance that determines the possibility of transition to the one-liquid description for the given problem.

\section{ANALOGY WITH THE DYNAMICS OF LIQUID HELIUM IN AN ELECTRIC FIELD}
\label{sec:Helimumlectric}

Let us demonstrate that Eqs.  (\ref{system1})--(\ref{system4}) are
identical to the equations appearing in the description of instability of
the electron-charged free boundary of liquid helium in an external
electric field. We assume that liquid helium is at a low temperature so
that the normal phase is absent  ($\rho _n \approx 0$ and $\rho \approx
\rho _s )$. The unperturbed (planar) boundary is charged by electrons with
surface charge density $\sigma $. It is well known that electrons can
freely move over the boundary, thus ensuring that the boundary is
equipotential one  \cite{EdelmanUFN1980,Shikin_UFN_2011}. In an applied
vertical uniform electric field, the field strengths over  ($E_o )$ and
inside ($E_i )$ the liquid are connected by relation $E_o - E_i = 4\pi
\sigma $.

Velocity potential $\Phi $ of the (single) liquid and electric field potentials over  ($\varphi_o$) and inside ($\varphi_i$) the liquid satisfy Laplace equations
\[
\nabla ^2\Phi = 0,\quad \quad \nabla ^2\varphi _{i,o} = 0.
\]
These equations must be solved together with the following conditions in the bulk and on the free boundary:
\begin{eqnarray*}
\Phi \to 0, \quad \varphi _{i,o} \to - E_{i,o} y, \qquad y \to -
\infty ,
\\
\rho \left( {\frac{\partial \Phi }{\partial t} + \frac{(\nabla \Phi )^2}{2}}
\right) = - \rho g\eta + \frac{\alpha \eta _{xx} }{(1 + \eta _x^2 )^{3 / 2}}
\\
+ \gamma - \frac{(\nabla \varphi _i)^2 -(\nabla \varphi _o )^2}{8\pi},
\qquad y = \eta ,
\\
\frac{\eta _t }{\sqrt {1 + \eta _x^2 } } = \partial _n \Phi , \quad
\varphi _i = \varphi _o = 0, \qquad y = \eta ,
\end{eqnarray*}
where the Bernoulli constant is $\gamma = (E_i^2 - E_o^2)/8\pi$, and the potential of boundary $y=\eta$ is assumed to be zero. The last term in the time-dependent Bernoulli equation is responsible for the electrostatic pressure at the free boundary; it includes the pressures over and under the surface.

The above equations turn out to be identical to Eqs.
(\ref{system1})--(\ref{system4}) for the quantum KHI (derived above in a
particular case when  $E_o = 0$ and, accordingly,  $\varphi_o \equiv 0$
and $E_i = - 4\pi \sigma$. This case (when the surface charge completely
screens the field over the liquid) was realized, for example, in
experiments \cite{VolodinHaikinEdelmanJETPLett1977,EdelmanUFN1980}. In the
framework of this analogy, auxiliary flow function  $\psi$ and electric
field potential  $\varphi_i$ in the liquid, as well as velocity difference
$V$ and field strength $E_i$ are connected by relations
\[
\psi \sqrt {4\pi \rho } \equiv \varphi _i ,\qquad
V\sqrt {4\pi \rho _n \rho _s / \rho } \equiv E_i.
\]
It should be noted that the Bernoulli constants also coincide in this case ($\Gamma \equiv \gamma$).

The revealed analogy makes it possible to use the results obtained earlier
from analysis of the electrohydrodynamic instability of the charged
surface of liquid helium
\cite{Gorkov_Chernikova_FNT_1976,Gorkov_Chernikova_DAN_1976,Zubarev_JETPLett_2000,Zubarev_JETP_2002,Zubarev_JETP_2008},
for analyzing the quantum Kelvin-Helmholtz instability.

Concluding this section, we note that this analogy cannot be extended to
the general 3D case. In the case of KHI, there exists a preferred
direction, viz., the direction of flow of the liquids  ($x$-axis in our
case). There is no preferred direction for electrohydrodynamic instability
because the problem is invariant to rotation about the vertical $y$ axis
along which the external electric field is directed.

\section{AMPLITUDE EQUATION FOR THE DYNAMICS OF THE FREE BOUNDARY}
\label{sec:amplitudeeq}

Analysis performed in \cite{Gorkov_Chernikova_FNT_1976} revealed that the
liquid helium boundary in an electric field becomes unstable when the
following condition holds: $E_i^2 + E_o^2 > E_c^2 $, where $E_c^2 = 8\pi
\sqrt {\rho g\alpha}$. In the vicinity of the instability threshold,
harmonics with wavenumbers close to $k_0 = \sqrt{\rho g /\alpha }$ grow.
Let us introduce the supercriticality parameter as
\[
\delta = \frac{E_i^2 + E_o^2 - E_c^2 }{E_c^2 }.
\]
If $|\delta| \ll 1$, it is natural to construct the equation for the
envelope for describing the dynamics of the boundary. For the 2D case (in
the 3D case, it is necessary to consider the interaction of three plane
waves with wavevectors turned through $2\pi / 3$
\cite{KuznetsovSpektorJETP1976,ZubarevZubarevaPhysLettA2000}), the shape of
the boundary is sought in form
\[
\eta (x,t) = \frac{1}{2k_0 }\left[ u(x,t)e^{ik_0 x} + u^*(x,t)e^{- ik_0} \right] + O\left( | u |^2 \right),
\]
where $u$ is the dimensionless complex amplitude (envelope) of the wave,
and asterisk marks complex conjugation. It was shown in
\cite{Gorkov_Chernikova_DAN_1976}, that the evolution of the boundary is
described by the Klein-Gordon equation with cubic nonlinearity:
\[
\frac{1}{2gk_0 }\frac{\partial ^2u}{\partial t^2} = \delta u +
\frac{1}{2k_0^2 }\frac{\partial ^2u}{\partial x^2} + \left( {\Delta ^2 -
\frac{5}{16}} \right)u\left| u \right|^2,
\]
where we have introduced notation
\[
\Delta = \frac{E_i^2 - E_o^2 }{E_c^2 }.
\]
It can be seen that in the linear approximation (in the spatially homogeneous case), the amplitude increases exponentially for $\delta > 0$. In this case, the nonlinearity hampers the instability development for  $0 < \Delta ^2<5/16$ and accelerates it for $\Delta^2> 5/16$.

In the case of our interest (when $E_o = 0$), for small supercriticality $\delta \approx 0$, we have $E_i \approx E_c $ and, hence, $\Delta \approx 1$. In such a case, the amplitude equation takes the form
\begin{equation}\label{key}
\frac{1}{2gk_0 }\frac{\partial ^2u}{\partial t^2} = \delta u +
\frac{1}{2k_0^2 }\frac{\partial ^2u}{\partial x^2} + \frac{11}{16}u\left| u
\right|^2,
\end{equation}
i.e., the nonlinearity is destabilizing. Analysis performed in Section~\ref{sec:Helimumlectric} shows that this equation also describes the evolution of the KHI of the He-II boundary. In this case, supercriticality is given by
\begin{equation}\label{porog}
\delta = \frac{V - V_c }{V_c },
\end{equation}
where the critical velocity is defined by Eq.\e{Vcdef}.

Due to the destabilizing effect of the nonlinearity, a tendency to an explosive increase in the amplitude of the He-II boundary appears during the KHI evolution in the spatially homogeneous solution; it increases unlimitedly over a finite time as
\[
u \propto \frac{1}{t - t_c }, \qquad t \to t_c ,
\]
where $t_c $ is the time of ``explosion.'' It is interesting that this
result is independent of the ratio of the densities of the normal and
superfluid helium components. The coefficient of the nonlinear term in Eq.
(\ref{key}) turns out to be universal. In the model developed here, the
only quantity depending on the ratio of densities is the difference in
velocities of the components  (\ref{Vcdef}), which is a threshold for the
KHI development and appears in supercriticality condition  (\ref{porog}).

After scaling
\[
t \to \frac{t}{\sqrt {2gk_0 } },\quad \quad x \to \frac{x}{\sqrt 2 k_0
},\quad \quad u \to \frac{4u}{\sqrt {11} }
\]
amplitude equation  (\ref{key}) for envelope $u$ assumes the following compact form:
\begin{equation}\label{KleinGordon1}
\frac{\partial ^2u}{\partial t^2}
=\delta u + \frac{\partial^2u}{\partial x^2} +\left| u \right|^2u.
\end{equation}
It corresponds to Hamiltonian
\begin{equation} \label{Hdimlessdef}
H = \int \left(
\left| \frac{\partial u}{\partial t} \right|^2
+ \left| \frac{\partial u}{\partial x} \right|^2
- \delta \left| u \right|^2
- \frac{\left| u \right|^4}{2} \right)dx,
\end{equation}
which is an integral of motion.

\section{CONDITIONS FOR EXPLOSIVE
DEVELOPMENT OF THE QUANTUM KELVIN-HELMHOLTZ INSTABILITY}
\label{sec:sufficient}

Thus, we have established that during the development of quantum KHI, wave
packet envelope $u$ obeys the $|\phi |^4$ Klein-Gordon complex nonlinear
equation with cubic nonlinearity. It is important that the
nonlinearity does not stabilize linear instability of Eq.
(\ref{KleinGordon1}), but on the contrary, enhances it, leading to an
explosive increase in amplitudes under certain conditions. Indeed,
assuming that perturbation of boundary $\eta(x,t)$ is localized in space
and, respectively, the amplitude $u(x,t)$ is localized in space, we
consider, analogously to Ref. \cite{KuznetsovLushnikov1995}, the temporal
evolution of square of the $L^2$ norm
\begin{equation}\label{Bdef}
B(t)\equiv \int {\left| u \right|^2dx}.
\end{equation}
Eqs. \e{KleinGordon1}--\e{Bdef} allow us to write that
    \begin{eqnarray} \label{Btteq}
     &B_{tt}=\int\left[2|u_t|^2+u_{tt}u^*+uu^*_{tt}\right]dx
\qquad\qquad \nonumber \\
     &\quad=-4 H+\int \left [ 6|u_t|^2-2\delta |u|^2+2|u_x|^2\right ]dx, \end{eqnarray}
where we have used integration by parts with respect to $x$ with allowance
for decreasing boundary conditions for $|x|\to\infty$. It should be
noted that contribution from  $2\int |u|^4dx$ was completely absorbed by
term $-4H$. The subscripts in Eq. \e{Btteq} and below indicate
differentiation: $u_t={\p u}/{\p t}$, $u_x={\p u}/{\p x}$, $B_{tt}={\p
^2B}/{\p t^2=d^2 B/d t^2}$, etc.

To obtain the estimate from below of the term
    \begin{eqnarray} \label{ut2}
     \int 6|u_t|^2dx=6\int R_t^2dx+6\int \phi_t^2R^2dx \end{eqnarray}
in Eq. \e{Btteq}, we write complex amplitude $u$ in form   $u\equiv R
e^{i\phi}$, where $R=|u|$ is the amplitude and $\phi$ is the phase. Using
the Cauchy-Bunyakovsky inequality $\left|\int fp\,dx \right|\,\le \left
(\int |f|^2 dx\right )^{1/2}\left (\int |g|^2 dx \right)^{1/2}$, which is
valid for complex-valued functions $f$ and $g$, we obtain inequalities
\begin{equation} \label{Rtinequal}
\left|B_t\right|=2\left|\int R R_t dx\right|\le2 B^{1/2} \left (\int R_t^2 dx\right )^{1/2}
\end{equation} and
\begin{equation} \label{Qinequal}
     |Q|=2\left|\int\phi_tR^2dx\right|\le 2 B^{1/2} \left (\int \phi_t^2R^2 dx\right )^{1/2},
\end{equation}
where $Q\equiv i\int[u_tu^*-u u^*_t]dx$ is the integral of motion
($Q_t\equiv 0$) of Eq. \e{KleinGordon1}. When Eq. \e{KleinGordon1} is used
in the quantum field theory and the theory of solitons, this integral is
sometimes referred to as the charge (see, for example,
\cite{BogolyubovShirkov1993,MakhankovPhysRep1978})); however, we will not
use this term below since the concept of charge has already been used in
Sections 4 and 5 in another context.

Using inequalities \e{Rtinequal} and \e{Qinequal} 29), we obtain the
following inequality from (\ref{ut2}):
\[
\int 6|u_t|^2dx\geq \frac{3B_t^2}{2B}+\frac{3Q^2}{2B}.
\]
Substituting this expression into (\ref{Btteq}) and omitting term $\int
2|u_x|^2dx$ (the disregard of this nonnegative term is compatible with the
sign of the inequality), we arrive at differential inequality
    \begin{equation} \label{43}
      B_{tt}\geq \frac{3}{2}\frac{B^{2}_{t}}{B}+\frac{3}{2}\frac{Q^2}{B} -4H-2\delta B.
    \end{equation}

Change of variables $B=A^{-2}$ allows us to write inequality \e{43} in the
form
\begin{equation} \label{45} A_{tt} \leq -\frac{\partial
U(A)}{\partial A}, \quad
\end{equation}
where
\begin{equation} \label{Upot}  \quad U(A)= -\frac {HA^4}{2}-\delta \frac
{A^2}{2}+\frac{A^6 Q^2}{8}.
\end{equation}

Differential inequality (\ref{45}) can be written in the equivalent form
of ordinary differential equation
    \begin{equation} \label{43b}
     A_{tt} = -\frac{\partial
U(A)}{\partial A}-h^2(t),
    \end{equation}
where $-h^2(t)$ is an unknown nonpositive force.

\begin{figure}
\includegraphics[width=0.49859\textwidth]{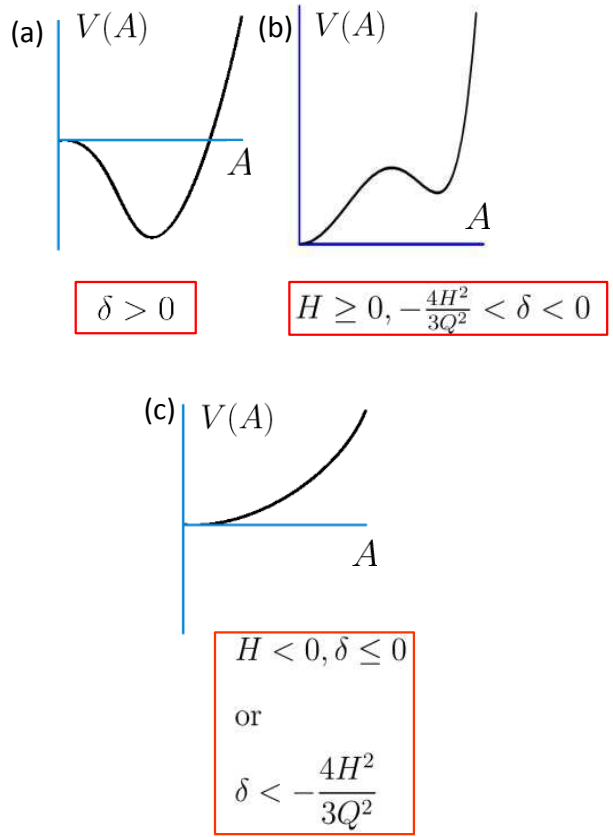}
\caption{
Qualitative form of potential $U(A)$ \e{Upot} as a
function of $H$ and $\delta$ for $Q\ne 0.$} \label{fig:schematic1}
\end{figure}

        Analysis of the formation of singularity in Eq.  \e{KleinGordon1}
        can be performed based on the method proposed in
\cite{LushnikovJETPLett1995} (see
\cite{LushnikovSaffmanPRE2000,LushnikovPRA2002,LushnikovPRA2010} for the
further development of this method). The method is based on analogy of Eq.
(33) with the equation of motion of an effective Newtonian ``particle''
with coordinate $A$  in potential  \e{Upot} under the action of an
additional (generally, nonpotential) force $-h(t)^2$, which pulls the
particle to the origin of coordinates. When this particle achieves zero
$A=0$, singularity $B=\infty$ is formed in Eq. \e{KleinGordon1}. The form
of the potential $U(A)$ \e{Upot} is shown qualitatively in Fig.
\ref{fig:schematic1} for $Q\ne 0$ depending on values of $H$ and $\delta$.
(The particular case $Q=0$ can be considered analogously; see also
\cite{KuznetsovLushnikov1995}).

It is convenient to introduce particle energy
\begin{equation}
\label{Wdef1} W(t) \equiv \frac{A_t^2}{2}+U(A),
\end{equation}
which depends on time due to the presence of force $-h(t)^2$ as
\begin{equation} \label{Wt} \frac{dW(t)}{dt} =A_t\left[A_{tt} +\frac{\p
U(A)}{\p A}\right ]=-h(t)^2 A_t .
\end{equation}

The complete classification of sufficient conditions for the formation of
the singularity over a finite time (also known as the wave collapse or
just collapse \cite{KuznetsovZakharov2007}) can be obtained. The
corresponding exact theorem can easily be formulated (when needed) based
on the following considerations that should be analyzed separaterly for
$A_t(0)>0$ and $A_t(0)\le 0$ as follows:

(A) If for the given initial conditions $A(0)$ and $A_t(0)\le0$ the
particle reached the origin ($A=0$ in Eq. \e{43b}) under the action of
conservative force $-\frac{\partial U(A)}{\partial A}$ only, then it would
definitely reach the origin of coordinates in the same or shorter time if
force $-h(t)^2$ were taken into account. This is due to the fact that in
accordance with relation \e{Wt}, in the case with $A_t\leq 0$ considered
here, we have inequality $W(t)\ge W(0)$ for the energy.  In the case
depicted in Fig. \ref{fig:schematic1}a, collapse occurs when  $W(0)>0$,
i.e., when the particle has sufficient initial energy for reaching zero
during a finite time. In the case shown in Fig. \ref{fig:schematic1}b, it
is necessary that either $A(0)$ be on the left of the barrier (for any
$W(0)$), or the value of $W(0)$ be over the barrier (for $A(0)$ on the
right of the barrier). The case illustrated in Fig. \ref{fig:schematic1}c
is the simplest because the collapse occurs here for any values of $A(0), \ A_t(0)$, and $W(0).$

(B) For $A_t(0)>0$, the sufficient conditions for the collapse can be
formulated for the cases shown in Figs. \ref{fig:schematic1}b and
\ref{fig:schematic1}c. In the case illustrated in Fig.
\ref{fig:schematic1}c, the collapse occurs for any values of $A(0), \
A_t(0)$ and $W(0)$, because the monotonicity of potential $U(A)$ stops
the motion of the particle to the right over a finite time (nonzero force
$-h(t)^2$ only accelerated this process), after which the particle falls
to zero during a finite time (in this case also, nonzero force $-h(t)^2$
only accelerates this process). To ensure, for example, the fulfillment of
condition $H<0$ for $\delta<0$, it is necessary in this case that
nonlinearity be quite strong for the negative contribution from term
$-\int \frac{1}{2}|u|^4dx$ in the Hamiltonian \e{Hdimlessdef} to exceed
the contributions from all remaining positive terms. In the case shown in
Fig. \ref{fig:schematic1}b with
 $A_t(0)>0$, the
collapse appears a fortiori if $A(0)$ is on the left of the barrier, and
the initial energy  $W(0)$   is insufficient for overcoming the barrier
even when force $-h^2(t)$ is ignored. With allowance for force $-h^2(t)$,
the particle necessarily stops on the left of the barrier and then falls
to zero over a finite time. In the remaining cases shown in Figs.
\ref{fig:schematic1}a and \ref{fig:schematic1}b ($A(0)$ is on the right of
the barrier), the particle can stuck in the vicinity of the potential
minimum because, as follows from Eq. \e{Wt} for $A_t(0)>0,$ we have $W(t)\le W(0)$; i.e., the particle loses energy, and after the reflection
from the wall, the energy may turn out to be insufficient for overcoming
the barrier. For this reason, the sufficient condition for the collapse in
these case cannot be formulated (although collapse is still possible, but
its full description requires detailed knowledge of the $-h(t)^2$
dependence).

If one of the above sufficient conditions for the explosive KHI evolution
holds and $A_t(0)\le 0$, the collapse time   $t_c$ for the formation of
the singularity satisfies the following inequality
\[
t_c \le \int\limits_0^{A(0)} {\frac{dA}{\sqrt {2[W(0)-U(A)]}}},
\]
which follows from Eqs.  \e{43b} and \e{Wdef1}.

It should also be noted that for spatially homogeneous initial conditions
$-h^2(t)\equiv 0$ (all integrals in this case must be considered in the
sense of their values per unit length along the $x$ axis), all
inequalities of this section become equalities; among other things,
inequality \e{45} becomes an ordinary differential equation for a
Newtonian particle. Therefore, the sufficient criteria for the collapse on
the class of homogeneous solution in this section become sufficient and
necessary conditions for the collapse, which generalizes the criteria for
collapse from Ref. \cite{KuznetsovLushnikov1995}, where the
contribution from integral of motion $Q$ was not taken into account. The
asymptotic form of falling of the particle to the zero $A=0$ corresponds
to constant velocity $A_t$; therefore, for $B=A^{-2}$ and, accordingly, for
squared amplitude $|u|^2$  of the envelope of a surface wave, the
asymptotic dynamics of the collapse corresponds to the law $(t_c-t)^{-2}$.

\section{Conclusions}
\label{sec:conclusion}

It should be noted that some of sufficient criteria for the explosive
increase of the amplitudes, which were formulated in Section
\ref{sec:sufficient}, are applicable in the case when a plane surface is
stable to small perturbations ($\delta < 0$).  This means that the
excitation of instability is hard, and a large initial perturbation of a
linearly stable regime may lead to the emergence of a singularity over a
finite time. In all cases, Eq.~(\ref{KleinGordon1}) in the vicinity of the
singularity becomes inapplicable: next orders of perturbation theory now
make a contribution of the same order of magnitude as the nonlinearity in
(\ref{KleinGordon1}). It can then be concluded that the solution
becomes strongly nonlinear in finite time $t_c$, i.e., the characteristic
slopes of the surface become of the order of unity due to the action of
the leading nonlinearity of the Klein-Gordon equation. After this, we
generally expect the breaking of waves. In this region, it is necessary to
consider complete hydrodynamic equations of Section \ref{sec:singlfluid},
which is beyond the scope of this article.

It should be noted that indefinitely strongly nonlinear stages of the
quantum KHI were analyzed in \cite{LushnikovZubarevPRL2018} disregarding
gravity and capillarity; complete integrability of the equations of motion
was demonstrated in the sense of reduction of exact dynamics to the
Laplace growth equation that has an infinitely large number of integrals
of motion and is associated with the dispersionless limit of the Toda
hierarchy \cite{MineevWeinsteinWiegmannZabrodinPRL2000}. Analysis of the
possible integrability of complete hydrodynamic equations of Section
\ref{sec:singlfluid} is also an interesting subject for future
investigations.

{\it Acknowledgements} The work of N.M.Z was partially supported by the
Russian Academy of Sciences (program no. 2 of the Presidium of the Russian
Academy of Sciences), Ural Branch of the Russian Academy of Sciences
(project no. 18-2-2-15), and Russian Foundation for Basic Research
(project no. 19-08-00098). The research of P.M.L. on analysis of
explosive instability was performed within State assignment ``Dynamics of
Complex Media.’’ The research of P.M.L. was supported by the National
Science Foundation (grant no. DMS-1814619). Simulations were performed  at
the Texas Advanced Computing Center using the Extreme Science and
Engineering Discovery Environment (XSEDE), supported by NSF Grant
ACI-1053575.



\end{document}